\documentclass[conference]{IEEEtran}
\makeatletter
\def\ps@headings{%
\def\@oddhead{\mbox{}\scriptsize\rightmark \hfil \thepage}%
\def\@evenhead{\scriptsize\thepage \hfil \leftmark\mbox{}}%
\def\@oddfoot{}%
\def\@evenfoot{}}
\makeatother
\usepackage[utf8]{inputenc}
\usepackage[T1]{fontenc}
\usepackage{fixltx2e}
\usepackage{flushend}
\usepackage{graphicx}
\usepackage{longtable}
\usepackage{float}
\usepackage{wrapfig}
\usepackage{soul}
\usepackage{textcomp}
\usepackage{marvosym}
\usepackage{wasysym}
\usepackage{latexsym}
\usepackage{amssymb}
\usepackage{hyperref}
\usepackage{url}
\usepackage{color}
\usepackage{verbatim}
\usepackage{mdwlist}

\title{Internet Topology over Time}
\author{
  Benjamin Edwards\\
  University of New Mexico\\
  bedwards@cs.unm.edu\\
  \and
  Steven Hofmeyr\\
  LBNL\\
  shofmeyr@lbl.gov
  \and 
  George Stelle\\
  University of New Mexico\\
  stelleg@cs.unm.edu
  \and
  Stephanie Forrest\\
  University of New Mexico\\
  forrest@cs.unm.edu\\
}

\date{}

\begin{document}

\maketitle

\newcommand{\fig}[2]{
 \begin{figure}[t]
    \centering
    \includegraphics[width=1.0\linewidth,clip]{#1.pdf}
    \vspace{-.3in}
    \caption{#2}
    \label{fig:#1}
  \end{figure}}

\begin{abstract}

  There are few studies that look closely at how the topology of the
  Internet evolves over time; most focus on snapshots taken at a
  particular point in time. In this paper, we investigate the
  evolution of the topology of the Autonomous Systems graph of the
  Internet, examining how eight commonly-used topological measures
  change from January 2002 to January 2010. We find that the
  distributions of most of the measures remain unchanged, except for
  average path length and clustering coefficient. The average path
  length has slowly and steadily increased since 2005 and the average
  clustering coefficient has steadily declined. We hypothesize that
  these changes are due to changes in peering policies as the Internet
  evolves. We also investigate a surprising feature, namely that the
  maximum degree has changed little, an aspect that cannot be captured
  without modeling link deletion. Our results suggest that evaluating
  models of the Internet graph by comparing steady-state generated
  topologies to snapshots of the real data is reasonable for many
  measures. However, accurately matching time-variant properties is
  more difficult, as we demonstrate by evaluating ten well-known
  models against the 2010 data.

\end{abstract}

\section{Introduction}

\label{sec:introduction}

The Internet is growing rapidly, having more than tripled in size in
the last decade, from 10,000 Autonomous Systems (ASes) in 2002 to
34,000 in 2010. However, few studies have looked carefully at the
time evolution of the Internet topology at the AS-level. Most studies
consider a snapshot of the AS-level topology of the Internet, derived
from the latest data available at the time of the research,
e.g.~\cite{Barabasi1999, RichClubUnnormalized, ClausetPL, ASIM}.

In this paper, we are interested in how the topology of the AS-level
Internet changes over time. We take the common approach of regarding
the Internet as a graph, where the vertices are ASes and the edges are
routing links between them. There are many properties of the Internet
graph that can be investigated, from the simple degree distribution to
more complex measures such as betweenness centrality. We select a set
of eight commonly used measures that are relevant to the way the
Internet functions. The measures are described in
\autoref{sec:measures}.

We investigate how the selected measures change over the period from
January 2002 to January 2010, and present the results in
\autoref{sec:results}. We find that the distributions for most of the
measures remain unchanged except average path length and clustering
coefficient. Since 2005, the average path length has slowly and
steadily increased and the average clustering coefficient has steadily
declined.  These results may signify changes in peering relationships
in the Internet; we discuss this idea in \autoref{sec:discussion}.

Our results imply that 
Internet topology models can be evaluated using single snapshots of
the topology in time for many measures but not all. We can expect
that models that matched invariant measures five years ago will still
match today. To this end we include a brief summary of 10 well-known
models and their performance on the latest data set in
\autoref{sec:ModelResults}. We find that models that were accurate
when originally proposed, often many years ago, still accurately
predict many time-invariant AS features (such as centrality), while
doing a poorer job on the time-variant measures, such as clustering
coefficient. In addition, the unchanging maximum degree of the
Internet is often poorly predicted. 

\section{Topological properties of the Internet}
\label{sec:measures}

We selected a set of eight measures for our analysis of Internet
topology evolution. Although many more measures are available,
e.g.,~\cite{Assortativity,Fractal,MeasureDump,3DataSourcesJointDegreeDistribution},
we chose those most commonly used in the past to evaluate both
generative models and data of the actual AS
topology~\cite{Haddadi2008}. Further, we selected measures that seem
most relevant to understanding the functioning of the Internet.
\autoref{tbl:MetJust} summarizes the measures and our rationales for
choosing them. Given space limitations, we restrict our attention to
graph-based measures,
ignoring network operation constraints, traffic flow analysis, and
distributions of AS relationship types.

\begin{table*}[htb]
\begin{center}
  \begin{tabular}{|l|l|}
    \hline
    Measure                              &  Rationale    \\
    \hline
    Degree Centrality                   &  Simplest way to measure AS prominence   \\
    Betweenness Centrality              &  AS prominence under best
    (shortest-path) routing. Inversely related to robustness to node deletion\\
    Page Rank Centrality                &  AS prominence under average (random)
    routing.     \\
    Path Length                         &  Related to routing efficiency (hops
    between source and destination)                   \\
    Clustering Coefficient              &  Related to the peering structure of
    the Internet, and routing resilience (number of alternative routes) \\
    K-Cores Decomposition               &  Related to tier structure of the AS Graph   \\
    Assortativity                       &  Relevant to peering relations \\
    S-Metric                            &  Distinguishes among scale-free graphs, alternate measure of assortativity \\
    \hline
  \end{tabular}
\end{center}
\vspace{-.05in}
\caption{Summary of measures.}
\label{tbl:MetJust}
\end{table*}

The measures are divided into four categories: \emph{Node Centrality},
\emph{Path Length}, \emph{Community Structure}, and \emph{Scale Free
  Structures}.  We also track simple properties such as the maximum
and average degree of ASes.

\subsection{Node Centrality}
\label{subsec:Centrality}

Node centrality measures are related to the prominence of ASes, which
is important when evaluating the implications of ISP
regulation~\cite{HofmeyrEtAl11} or the robustness of the
Internet~\cite{DoyleAldersonRobustFragile}. We use three measures of
centrality: the node \emph{degree}, the \emph{betweenness}, which is
the fraction of all shortest paths that pass through a
node~\cite{Freeman77}, and the \emph{page rank}, which is the number
of times that a node will be visited on a sufficiently long random
walk on the graph~\cite{pagerank}. The betweenness centrality is
inversely related to the robustness of the graph to removal of nodes,
because the more paths that pass through a node, the more damage will
be done when that node is removed.

\subsection{Path Length}
\label{subsec:PathLength}

The \emph{average shortest path length} from a node to all other nodes
in the graph (the geodesic distance)\footnote{This is sometimes
  referred to as \emph{closeness centrality}, though there are other
  definitions for closeness centrality~\cite{Closeness}.} is important
because it relates to the number of routing hops between ASes. Not all
packets travel along the shortest paths because of business agreements
(such as the valley-free rule), but to a first approximation, routing
distances (hops on the AS graph) are largely determined by the
shortest paths.

Alternative path length measurements include diameter, which is the
longest of the shortest paths between all pairs of nodes, and the
effective diameter, which is the path length that defines the 90th
percentile of all paths~\cite{Leskovec2007}. In the Internet, the
distribution of path lengths has small variance, so diameter and
effective diameter are only slightly larger than the shortest average
path length and highly correlated with it. Hence we use only the
average shortest path length.

\subsection{Community Structure}
\label{subsec:Community}

Community structure measures  
how groups of nodes form substructures within the graph and is
relevant to understanding various aspects of the Internet, such as the
tiered structure and resilience to node deletions.  Although there are
many community structure measures
\cite{FreemanCommunity,CommunityDetections}, we chose three that 
reveal important features of the AS graph. The first measure is the
\emph{local clustering coefficient}, which is the number of edges
among the neighbors of a node as compared to the maximum possible
number~\cite{Clustering}.\footnote{We do not use transitivity, which
  is an alternative definition of the clustering coefficient, because
  it tends to be highly correlated with the average degree and so does
  not yield additional useful information.} The clustering coefficient
is related to the resilience of the routing infrastructure, because it
reflects the number of alternative routes between pairs of nodes (for
example, a tree has a coefficient of 0 and the removal of any edges
will partition the graph). The second community structure measure is
\emph{degree assortativity}, which measures whether nodes tend to
connect to others of similar degree~\cite{Newman03}. The final measure
is \emph{k-cores decomposition}, which measures successive maximally
connected subgraphs~\cite{kCores}. We report two measures for k-cores:
the distribution of k-cores values, and the size of the maximum core,
k-max.

\subsection{Scale-free Structures}
\label{subsec:Smetric}

The power-law degree distribution of the AS graph is a scale-free
property often cited as a distinguishing feature of the Internet,
e.g.,~\cite{ClausetPL}. If the AS graph can be described as scale-free
it may share properties with other scale-free networks, for example,
the tendency to be `robust yet fragile' or the preferential attachment
growth dynamic. However, Li et al~\cite{li2005towards} showed that it
is possible to construct graphs and general data sets that have
similar scale-free properties but very different structures. To
address this issue, Li et al. propose the \emph{s-metric}---the sum
over edges of the product of the degree of the two nodes an edge
connects. This computation yields a single value that measures the
extent to which a graph is actually scale-free.

\section{Data Sets} \label{sec:data}

To investigate how the Internet changes over time, we collected a set
of AS graphs covering the period from January 2002 to January 2010 by
parsing monthly snapshots of BGP routing table dumps from Oregon Route
Views and RIPE.\footnote{\url{www.routeviews.org}
  and\url{www.ripe.net}} Although BGP routing tables are dumped every
few hours, monthly snapshots were of sufficient temporal resolution
given the long time scale of the analysis. The monthly snapshots were
compiled by parsing all of the dumps from the first day of each month,
taking every adjacent pair in the ASPATH and adding them to the graph
for that month. We did not filter out self loops, private Autonomous
Systems Networks (ASNs), or any other potential spurious or inaccurate
results from the dumps, as it is generally assumed that the number of
false positives of this type are small~\cite{Bigfoot}.

One potential problem with the BGP data is that nodes and edges
disappear and reappear due to the way the data are sampled. While
there are other ways of dealing with disappearing edges and
nodes~\cite{Leskovec2007}, we assumed that nodes and edges that
temporarily disappear from the BGP tables actually exist throughout
from first appearance to last.  Our data set is
available\footnote{\url{https://ftg.lbl.gov/projects/asim/data-2/}} in
networkx\footnote{\url{http://networkx.lanl.gov/}} format.

To validate our results, we ran identical experiments using data
collected from the Cooperative Association of Internet Data Analysis
(CAIDA)~\cite{caida}, and obtained essentially identical results. The
collectors of the CAIDA data sets went to great length to deal with
various false-positive errors and the issue of nodes and edges that
temporarily disappear, so the close agreement between the two data
sets indicates that relatively simple preprocessing of the data is
adequate for the purposes of our study.

Although false positives in the data are likely rare, false negatives
(missing links) are likely common because the BGP dumps do not capture
peering links between smaller ASes on the edge of the
graph.\footnote{Roughan et al~\cite{Bigfoot} estimate that an AS graph
  extracted from public BGP views is likely to miss 27\% of links
  overall, and 70\% of peer-to-peer links.}  Although several studies
have attempted to quantify the number of missing links
(e.g.~\cite{Bigfoot,Cohen06theinternet,HeEtAl2009}), it is difficult
to determine how exactly those hidden links could affect the structure
of the AS-graph. Consequently, we focus on the \emph{visible}
Internet, in which we see subtle topological changes
(see~\autoref{sec:results}) that we speculate could be caused by an
increase in missing links.

\section{Results} 
\label{sec:results}

Most of the measures yield a distribution rather than a single value.
Although we can plot the distributions together, year by year, it is
also useful to have a single value for determining the changes over
time. A common approach to this problem aggregates distributions,
using measures of central tendency, extent, or
spread~\cite{Haddadi2008}. However, studying the distributions as a
whole before aggregating allows us to discover changes
to the shape of the distribution (e.g. a transition from an
exponential distribution to a power-law) that might not be revealed
under aggregation. Consequently, we test whether the distributions
between years differ using the Cramér–von Mises Criterion (CMC)
\cite{Anderson1962}. The CMC tests the hypothesis that two samples of
data are drawn from the same distribution. Although many alternative
tests and measures exist~\cite{JSDivergence,KSStat,Rogers97}, the CMC
gives accurate comparisons and captures intuitive similarities between
plots that can be seen visually.

We used the CMC to identify year by year changes in all of the measures
that have distributions. \autoref{tbl:MetricChange} shows the changes
from one year to the next from 2002 to 2010. For each year, we applied
the measures to the AS-graph for June; varying the month of data
collection does not vary the results. For measures that do not have
distributions (k-max, assortativity and the s-metric),
\autoref{tbl:MetricChange} reports the absolute values, rather than
the year by year differences.

\begin{table*}
\begin{center}
\begin{tabular}{|l|l|l|l|l|l|l|l|l|l|l|l|l|l|}
  \hline
  Year & Nodes & Edges & Degree & Betweenness & Page Rank & Path Length &
  Clustering & K-Cores & K-max & Assort. & S-Metric \\
  \hline
  2002 & 13172 & 26695 & & & & & & & 16 & -0.189 & .0114\\
  \hline
  2003 & 15446 & 32089 & NS & NS & NS & **** & NS & NS & 18 & -0.188 & .0121\\
  \hline
  2004 & 17722 & 39654 & NS & NS & NS & **** & NS & NS & 22 & -0.188 & .0159\\
  \hline
  2005 & 20174 & 45505 & NS & NS & NS & **** & NS & NS & 24 & -0.196 & .0176\\
  \hline
  2006 & 22708 & 50796 & NS & NS & NS & **** & * & NS & 24 & -0.189 & .0183\\
  \hline
  2007 & 25691 & 58200 & NS & NS & NS & **** & NS & NS & 26 & -0.189 & .0199\\
  \hline
  2008 & 28640 & 64111 & NS & NS & NS & **** & ** & NS & 25 & -0.192 & .0203\\
  \hline
  2009 & 31645 & 69938 & NS & NS & NS & **** & * & NS & 25 & -0.195 & .0210\\
  \hline
  2010 & 34055 & 71544 & NS & NS & NS & **** & **** & NS & 21 & -0.191 & .0179\\
  \hline
\end{tabular}
\caption{Annual change of measures. We report computed values for k-max,
  assortativity and s-metric. For other measures, we report the statistical
  significance of the change between years (NS: Not Significant, *: $p<.1$, **:
  $p<.05$, ***: $p<.01$, **** : $p<.001$). }
\label{tbl:MetricChange}
\end{center}
\end{table*}

\subsection{Unchanging Features}

\autoref{tbl:MetricChange} shows that the node centrality measures
(degree, betweenness and page rank) stay constant over
time. \autoref{fig:DegreeDistByYear} illustrates this point, showing
that the power-law degree distribution is
virtually identical over time. We obtain similar results for
betweenness and page rank---the distributions are stable over time
(data not shown).

\fig{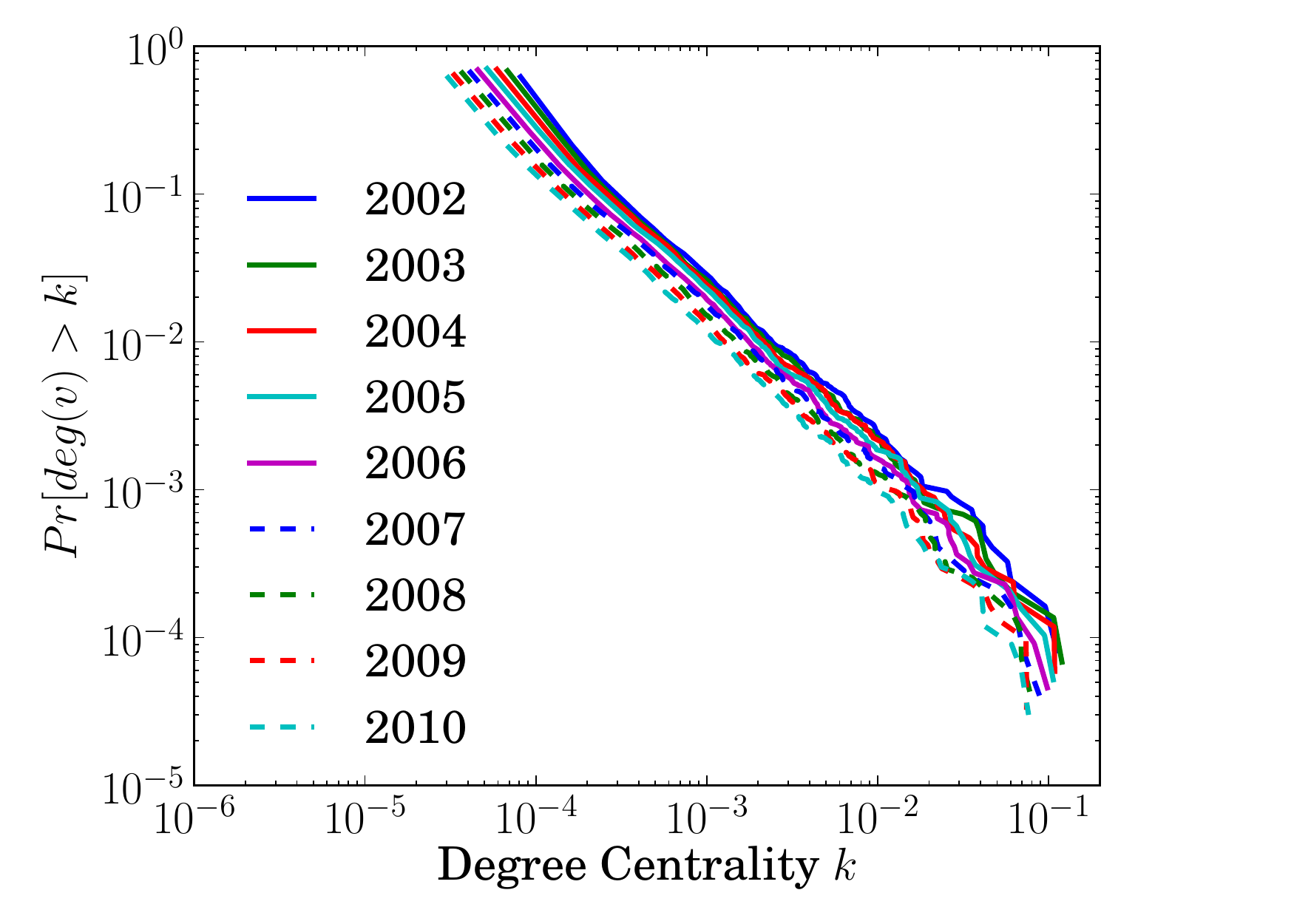}{Complement cumulative distribution of the
  degree of ASes between 2002-2010}

In \autoref{fig:DegreeDistByYear} not only is the slope of the
distribution unchanging, but the \emph{extent} (maximum degree) is
nearly constant. Only three ASes have had the maximum degree in the
years 2002 to 2010: MCI Inc., Level 3 Communications, and Cogent
Communications (see \autoref{fig:TopkASes}). MCI, which held the top
position until late 2008, maintained a nearly constant degree, while
Cogent and Level 3 had a nearly monotonic increase in degree over the
same time period. Growth is not guaranteed for an AS: 61\% of all ASes
experienced a month to month decline in degree at least once between
2002 and 2010 and fully 84\% of ASes that have existed for five or
more years experienced at least one monthly decline.

\fig{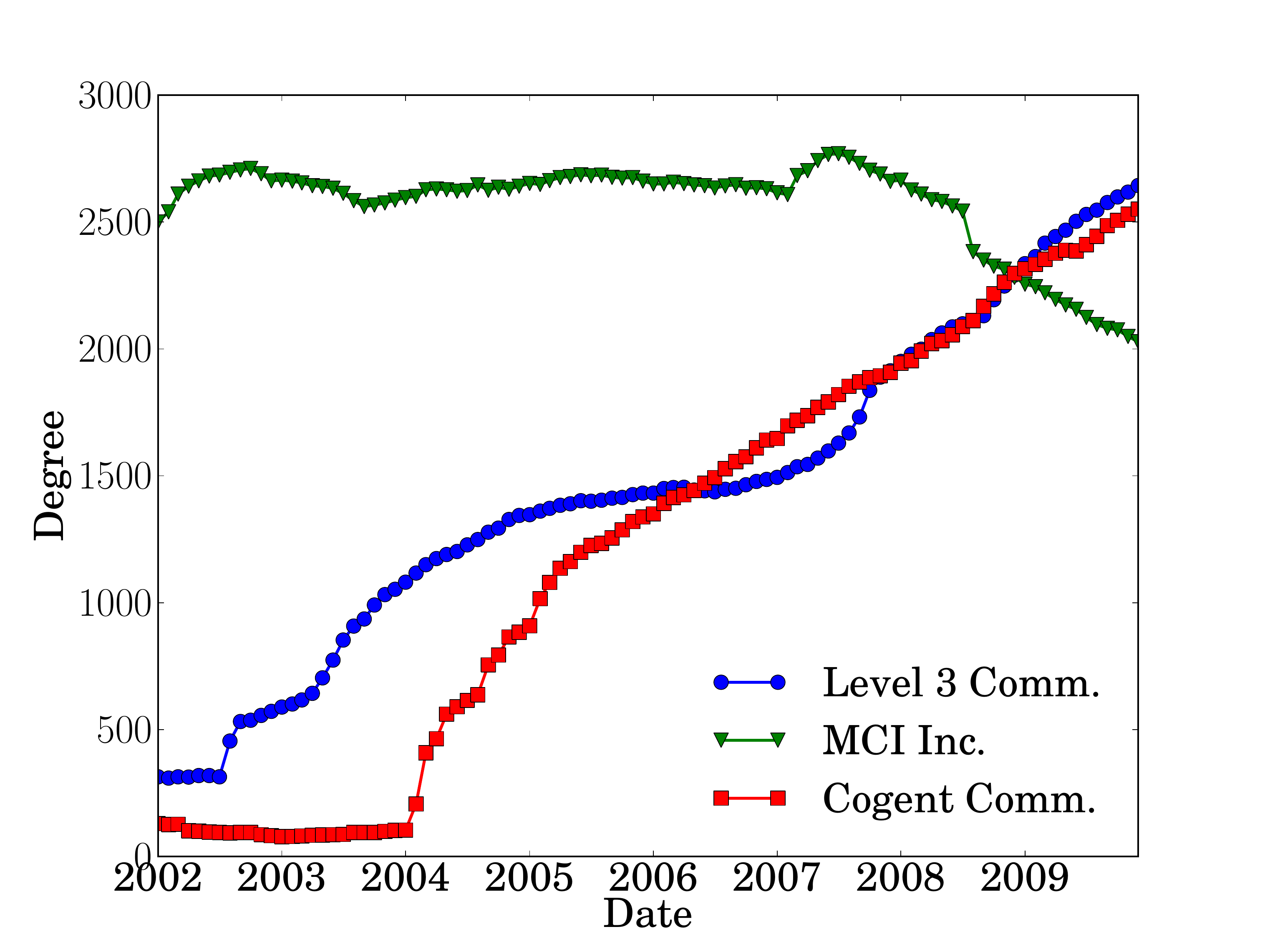}{Degree changes in the three largest degree ASes.}

Other clearly unchanging features in \autoref{tbl:MetricChange} are
the k-cores and the degree assortativity. The s-metric appears to
gradually increase over the entire period, but this could be a
meaningless change: According to Li et al.~\cite{li2005towards},
graphs with low s-metric values (below 0.1) are likely ``scale-rich''
and difficult to differentiate from each other using the s-metric.

\subsection{Changing Features} \label{subsec:change}

Two measures exhibit distinct changes over time: the average path length and the
clustering coefficient. The changes in the distribution of path length, although
statistically significant, are not easily visible on a plot. The clustering
coefficient, on the other hand, has changes that are clearly visible, as can be
seen in \autoref{fig:CCDistByYear}. Although the distribution changes little in
shape, there is a distinct downward trend over time.

\fig{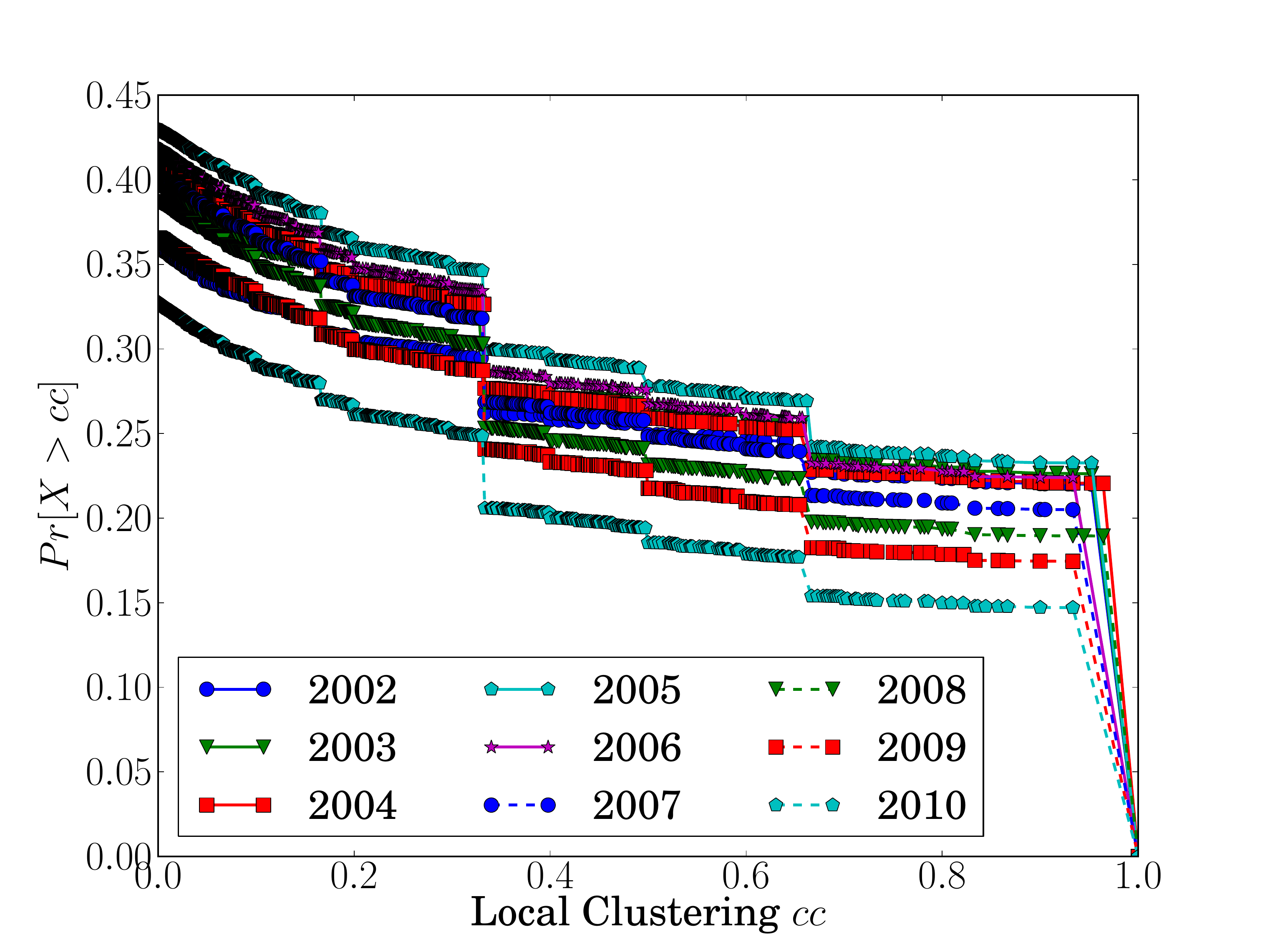}{Complement cumulative distribution of the local
  clustering coefficient of the AS graph}

\fig{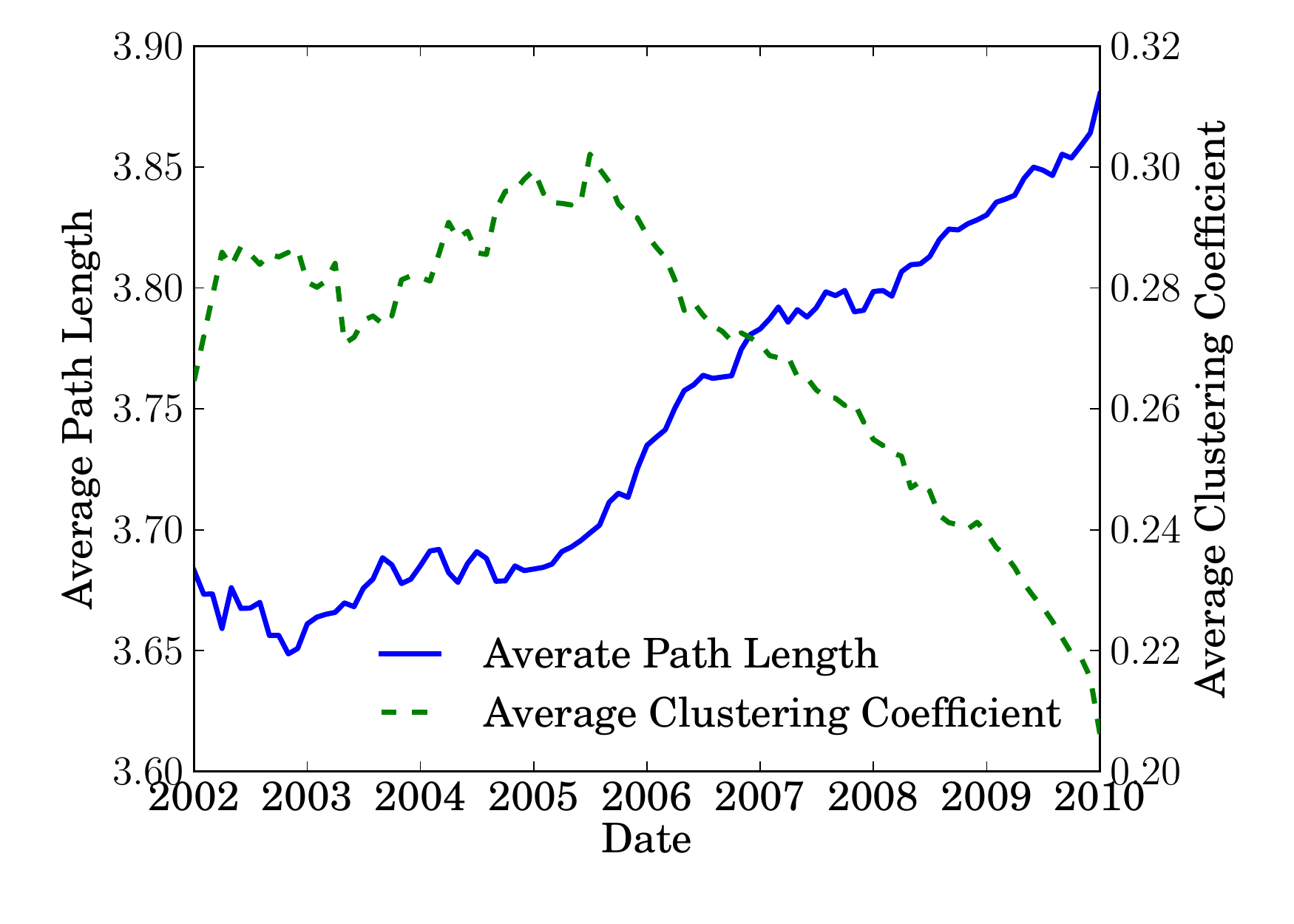}{The change in the average path length
  and the average clustering coefficient, plotted monthly. Both
  clustering coefficient and path length experience a nearly constant
  variance of .16 and .3 respectively, and so we do not report error
  bars in this figure.}

The trends in path length and clustering coefficient are evident in
\autoref{fig:AveragePathLengthCCByYear}, which plot averages over
time. We can see that around 2005, the structure of the visible AS
graph began to change. The average path length and average clustering
coefficient do not exhibit any clear trend until mid 2005, when they
began slowly increasing and decreasing respectively. In addition, the
k-max value was steadily increasing up to 2005, after which it leveled
off (see \autoref{tbl:MetricChange}). From these changes we can infer
that both the efficiency of the \emph{visible} Internet (as measured
by number of hops on the shortest paths between locations) and
resilience to router failure, have been decreasing since 2005. We
discuss possible reasons for these changes in
\autoref{sec:discussion}.

The shift in topology beginning in 2005 indicates that the Internet
has not reached steady-state. This observation contradicts many models
of the AS graph, in which measures converge to a steady-state
distribution. The next section studies how well different models of
Internet topology match the AS graph on the different measures.

\subsection{Topological Models of the Internet}
\label{sec:ModelResults}

The relatively slow changes we see in the Internet imply that
generative models that give accurate results at one size should do so
at most sizes. To check this we generated topologies from 10 different
models and compared them to the AS graph from June 2010.
\autoref{tbl:models} lists the models we considered. In each case, we
generated a topology of 34,055 nodes, the same number as our latest
snapshot of the AS graph.

\begin{table*} 
\begin{center}
 \begin{tabular}{|l|l|l|l|} \hline
Name & Abbr. & Description & Reference \\ \hline \hline
Bar\'abasi-Albert & BA & Original preferential attachment model &
\cite{Barabasi1999} \\ \hline Heuristically Optimized Trade-offs & FKP
& Local optimization model & \cite{FKP} \\ \hline Generalized Linear
Preference & GLP & Modified preferential attachment model &
\cite{GeneralizedLinearPreference} \\ \hline Univariate Heuristically
Optimized Trade-offs & UFKP & Modified local optimization model &
\cite{ChangFKP} \\ \hline Bivariate Heuristically Optimized Trade-offs
& BFKP & Modified local optimization model & \cite{ChangFKP} \\ \hline
Multivariate Heuristically Optimized Trade-offs & MFKP & Modified
local optimization model & \cite{ChangFKP} \\ \hline Interactive
Growth & IG & Modified preferential attachment model &
\cite{PositiveFeedbackPreference} \\ \hline Positive Feedback
Preference (1) & PFP1 & Modified IG model &
\cite{PositiveFeedbackPreference} \\ \hline Positive Feedback
Preference (2) & PFP1 & Modified PFP1 model &
\cite{PositiveFeedbackPreference} \\ \hline ASIM & ASIM & Agent based
topology generator & \cite{HofmeyrEtAl11} \\ \hline \end{tabular}
\vspace{-.05in} 
\caption{Summary of models evaluated.}
\label{tbl:models}
\end{center}
\end{table*}

\fig{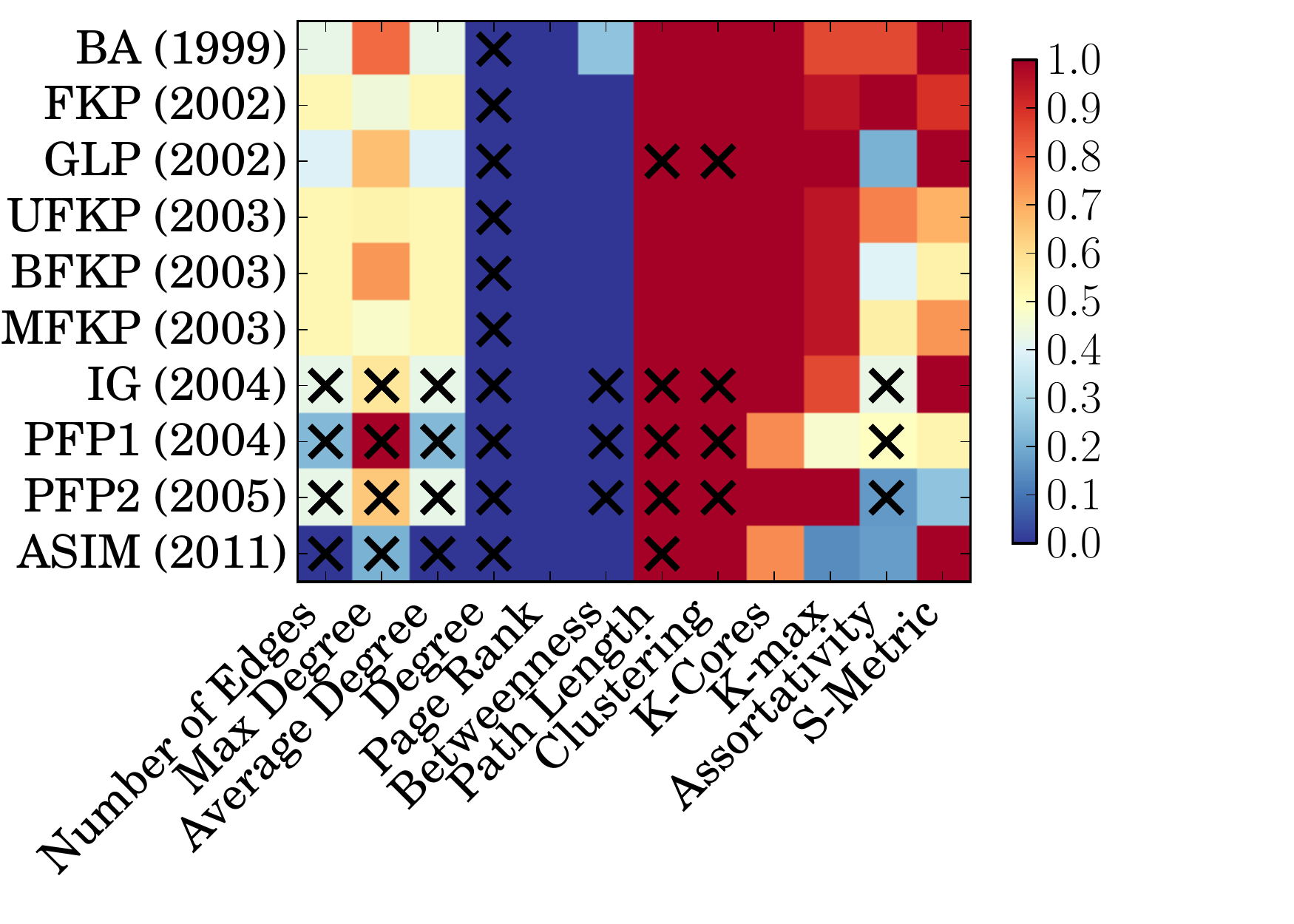}{Evaluating models. The color of each square
  corresponds to the statistical significance level for the
  Cramér-von Mises criterion when comparing the model to the January
  2010 AS graph for a given measure. The number of stars (0 to 4) is
  scaled between 0 and 1 for consistency with the other measures. For
  k-cores, degree assortativity and the s-metric, the color
  corresponds to the relative error. Hence lower values mean more
  similar distributions. X's indicate that the measure was reported in
  the original publication of the model.}

The comparisons of the different models to the AS graph are shown in
\autoref{fig:ComparisonChart}. As expected, later models are more
accurate than earlier models (the models are ordered from oldest at
the top to most recent at the bottom). All models generate close
matches to the degree distribution, which makes sense because this is
typically the first measure authors use to evaluate a model, and it has not
changed over time. Further, most models match the other measures of
centrality quite well, presumably because they are also
time-invariant.

It is notable that all the models perform worst on the measures that
change the most, namely path length and clustering. In general, models
match time-invariant measures better than time-varying measures, with
the exception of maximum degree, which no model captures well.

It is not surprising that most generative models we studied performed
well on the time-invariant measures the models were originally tested
on. However, most models perform much worse on measures they were not
originally tested against. This indicates the importance of evaluating
models against a wide variety of measures, because many different
graphs can be similar in one measure, but it is much harder for
different graphs to be similar in multiple measures.

\section{Related Work}
\label{sec:related}

There are few studies that look at changes in topological
characteristics beyond the number of nodes and edges. Most of those
that do focus on inter-AS relationships, for example, Chang et
al~\cite{ChangEtAl06} study the changes in customer-provider
relationships and find that the number of providers is increasing over
time. Another approach was taken by Oliveira et al~\cite{Oliveira06},
in which they investigate the changing relationship over time between
stub ASes and transit providers. They find that the net growth of
rewirings for transit providers levels off at the end of 2005, around
the same time our results show subtle changes in the Internet. Gill
et al~\cite{Gill2008} look more closely at the evolution of peering
relationships, and find that over time large content providers are
relying less on Tier-1 ISPs, and more on peering with lower
tiers. This finding is supported by Labovitz et
al~\cite{LabovitzEtAl2010}, who report a rapid increase in the traffic
flow over peer links over time, resulting in a less hierarchical
Internet topology. These observations could potentially explain some
of our results, as we discuss in \autoref{sec:discussion}.

In addition to studying business relationships, Dhamdhere el
al~\cite{Dhamdhere08} reported on changes in average degree and
average path length over time. Their results on path length agree with
ours, although their study included only data up to 2007, so the
trends are less clear. Because the degree distribution does not
change, it is likely that the shift they see in average degree is a
result of a steadily increasing sample size. Another study
\cite{Gaertler04} used spectral analysis to investigate
clustering on the AS graph, and study coreness and changing path
diversity. This analysis, however, covers short time spans (at most
two and a half years), and only considers data before 2004.

The work of Zhang et al~\cite{ASKCores} is perhaps closest to
ours. They study changes in several topological measures over the time
period from 2001 to 2006. Because of this time period, their results
do not capture the trends we report post-2005. However, the changes
they document agree with what we observed in the earlier period: They
find the assortativity and k-cores are stable over time and from
2004/2005 onwards, the k-max value changes little. Further, they find
the average clustering coefficient starts declining around 2005, and
the average path length starts increasing gradually.

\section{Discussion} 

\label{sec:discussion} 

We have reported a distinct shift in the topology of the \emph{visible} Internet
since 2005: the average path length is increasing, and the average clustering
coefficient is decreasing (\autoref{fig:AveragePathLengthCCByYear}). On the
surface, it would appear that the Internet is getting both less efficient and
less resilient. But this may not actually be the case, because the the shift is
likely caused by changes in peering policies that affect the hidden Internet and
cannot be measured with public BGP dumps. As mentioned in~\autoref{sec:related},
there are several studies showing that content providers are routing more
traffic over hidden peer-to-peer links, and relying less on the more publicly
visible Internet infrastructure. Consequently they have less need to establish
new customer-provider links, and a decreasing number of new customer-provider links increases the
\emph{observed} average path length of the graph.

Most models match topological measures that are invariant over time in
the AS graph, particularly centrality. However performance degrades
when examining time-variant measures such as average path length and
clustering coefficient. Future modeling efforts will need to focus on
incorporating mechanisms that can cope with such changing
dynamics. For example, few existing models allow for the loss of
links in the AS graph, a common feature according to our
data. Agent-based models such as ASIM are potentially a promising
direction for future AS topology modeling efforts because they can
naturally model economic pressures that lead to link
deletion. Further, robust statistical techniques such as the CMC will
be needed to verify topological results.

In conclusion, it is surprising that so few of the common measures of
Internet topology have changed over the past eight years, even though
the number of ASes has tripled during this time period. Those measures
that do change point to the increasing importance of understanding the
role of policy and economics in determining Internet topology. Going
forward, it will be increasingly important to find ways to reveal
hidden links and evolving peering relationships.

\section{Acknowledgments}
The authors gratefully acknowledge the support of DOE grant DE-AC02-05CH11231.
Stephanie Forrest acknowledges partial support of DARPA (P-1070-113237), NSF
(EF1038682,SHF0905236), and AFOSR (Fa9550-07-1-0532).

\bibliographystyle{IEEEtran}
\bibliography{IEEEabrv,paper}

\end{document}